\documentclass[10pt,letterpaper]{article}
\usepackage{opex3}

\begin{document}

\title{More studies on Metamaterials Mimicking de Sitter space}

\author{Miao Li, Rong-Xin Miao* and Yi Pang}

\address{Kavli Institute for Theoretical Physics, Key Laboratory
of Frontiers in Theoretical Physics, Institute of Theoretical
Physics, Chinese Academy of Sciences, Beijing 100190, People's
Republic of China}
\address{Interdisciplinary Center for Theoretical Study,
University of Science and Technology of China, Hefei, Anhui 230026,
China}
\email{*mrx11@mail.ustc.edu.cn} 


\begin{abstract}
We estimate the dominating frequencies contributing to the Casimir
energy in a cavity of meta- materials mimicking de Sitter space, by
solving the eigenvalue problem of Maxwell equations. It turns out
the dominating frequencies are the inverse of the size of the
cavity, and the degeneracy of these frequencies also explains our
previous result on the unusually large Casimir energy. Our result
suggests that carrying out the experiment in laboratory is possible
theoretically.
\end{abstract}



\section{Introduction}
The Casimir energy is one of the important predictions in quantum
field theory and continues to be source of inspiration for
theoretical as well as experimental work \cite{Casimir50}. It is the
regularized difference between two energies: one is the zero point
energy of the electromagnetic field in a finite cavity and the other
is that in an infinite background. Generically, the expression of
the Casimir energy depends on the details of the cavity including
properties of its bulk and its boundary. The earliest work
\cite{Casimir:1948dh} on the Casimir energy and the following up
studies
\cite{Boyer:1968uf,Balian:1977qr,Milton:1978sf,Plunien:1986ca,Bender:1976wb,Milton:1979yx,Milton:1983wy,Bordag:1996ma,Odintsov:1990qq}
all reported a result that the energy density of the Casimir energy
is inversely proportional to the forth power of the typical size of
the cavity. Applying this result to the universe, this kind of the
Casimir energy can not be taken as a possible origin of dark energy,
since $1/L^4$ is too small compared with the observed dark energy
density \cite{cc1,cc2} if $L$ is chosen to be the a typical size of
universe.

With doubt about the applicability of the previous results to de
Sitter space, we carried out a calculation on the Casimir energy of
electromagnetic field in the static patch of de Sitter space
carefully \cite{Li:2009pm}. We obtained a drastically different
Casimir energy which is proportional to the size of the event
horizon taking the same form as holographic dark energy
\cite{Li,Huang}. Then basing on the recent theoretical and
experimental development of metamaterials
\cite{Plebanski:1959ff,Leo,Pendry,Ge,
exp,Houck,Smith,Cubukcu,Yen,Linden,Zhang}, we proposed to design
metamaterials mimicking de Sitter (along the line of studying
cosmological phenomenon in laboratory, some other researches have
been carried out including the negative phase velocity of
electromagnetic wave in de Sitter, mimicking black hole and cosmic
string by metamaterials \cite{Mackay:2004ag,Cheng:2009ja,Mackay}).
Different from the usual ones, the permittivity and permeability
parameters both have divergent components tangent to the boundary.
This unusual fact leads to a brand new Casimir energy in the cavity
coinciding with the the result from the gravity side with the Planck
scale is replaced by some microscopic scale in metamaterials. We
encourage experimentalists to make such metamaterials and measure
the predicted Casimir energy. This work will have significant
implication to cosmology.

Apparently, it seems difficult to carry out such an experiment,
because the Casimir energy is expressed as a sum of all the
frequencies, meaning that each frequency contributes a part to this
energy, while metamaterials have frequency dispersion, in other
words, the designed permittivity and permeability is effective only
to frequencies in certain brand. However, this difficulty can be
circumvented as we will uncover a fact that there is a typical
frequency whose contribution to Casimir energy is dominating. Thus a
cavity of metamaterials effective at this typical frequency is
sufficient to mimic de Sitter space and induces the Casimir energy
predicted by theoretical calculation.

We shall show that the typical dominating frequency is $\omega\sim
1/L$, where $L$ is the size of the cavity, and the dominating
angular quantum number is $l\sim L/d$, $d$ is the short distance
cut-off. These two numbers conspire to give the Casimir energy
$L/d^2$, the same order of our previous result \cite{Li:2009pm}.

The rest of this paper is organized as follows. In sect.2, we
uncover the fact that the Casimir energy has a typical contributing
frequency. We estimate the typical frequency in metamaterials
mimicking de Sitter in sect.3. We conclude in sect.4.

\section{General discussion on the typical frequency of Casimir energy }

Physically, it is conceivable that there is a typical contributing
frequency to the Casimir energy. According to its definition, the
Casimir energy measures the difference between vacuum energy in a
finite cavity and that in infinite background. The former is usually
contributed by a discrete spectrum and the latter comes from a
continuous one. Thus the Casimir measures the difference between
discrete and continues ones. When frequency is large, the discrete
spectrum approaches a continues one, and their contributions cancel
with each other; for small frequencies, their contribution is also
negligible. Since the contribution from very large and very small
frequencies is tiny, there should be some intermediate scale at
which the difference between discrete spectrum and continuous
spectrum is maximum. Then the frequency at this scale is the typical
frequency.

As a heuristic example, we read off this typical frequency from the
process of computing Casimir energy in the static Einstein's
universe. In this case, the Casimir energy is given by
\cite{Ford:1975su}
\begin{equation}\label{CE1}
   E_c=\frac{1}{2a_0}\sum_{n=0}^{\infty}n^3-\frac{a^{3}_0}{2}\int_0^\infty
\omega^3d\omega,
\end{equation}
where the discrete spectrum consists of  $\omega=n/a_0$ with
degeneracy $n^2$ , and $a_0$ is the radius of Einstein universe.
Reparameterizing $\omega$ by $t/a_0$, then using Abel-Plana formula
\begin{eqnarray}\label{AP formula}
\sum^{\infty}_{n=0}F(n)-\int^{\infty}_{0}dt
F(t)=\frac{1}{2}F(0)+i\int^{\infty}_{0}dt\frac{F(it)-F(-it)}{e^{2\pi
t }-1},
\end{eqnarray}
eq.(\ref{CE1}) is equal to
\begin{equation}\label{E}
    E_{C}=\frac{1}{a_0}\int_0^\infty dt\frac{t^3}{e^{2\pi t
}-1}.
\end{equation}
This result is very interesting because it appears that Casimir
energy is contributed by frequencies satisfying a distribution
similar to blackbody. We read off from this formula that typical
frequency is at $t\approx 1/2\pi$ corresponding to
$\omega\approx1/(2\pi a_{0})$.
 However, in the general three
dimensional cases of physical interest, the discussion should take
into account complications due to at least two reasons. The first
one is that it is hard to find out the exact expression of the
discrete frequencies of electromagnetic wave in a finite cavity
which are eigenvalues of a complicated partial differential equation
deduced from Maxwell equations. The second is that in three
dimensions, one usually has two integers $(n,l)$ to denote the
discrete frequencies where $n$ is radial quantum number and $l$ is
angular quantum number, thus Abel-Plana formula should be utilized
repeatedly. Although it is difficult to derive an exact result,
estimations are always possible under some reasonable assumptions.
In the next section, we will try to solve the eigenvalue problem of
Maxwell equations in the metamaterials designed to mimic de Sitter
and estimate the typical frequency of the Casimir energy. Our
estimation is based on an assumption that the radial quantum number
$n$ is frozen then the Casimir energy comes mostly from frequencies
with smallest $n$. This is because in a finite cavity a frequency
usually grows linearly with $n$, thus its contribution is suppressed
exponentially by the blackbody factor. As a check of this
assumption, the large $n$ behavior of the eigenvalue will be given.
It is indeed a linear function of $n$.

\section{typical frequency of Casimir energy in metamaterial mimicking de Sitter}

The metamaterials mimicking de Sitter space is designed with the
following permittivity and permeability \cite{Li:2009pm}
\begin{equation}\label{M}
\varepsilon^{\tilde{r}\tilde{r}}=\mu^{\tilde{r}\tilde{r}}=L^{2}\frac{\sin^2(
\tilde{r}
/L)}{\cos(\tilde{r}/L)}\sin\theta,~~~~\varepsilon^{\theta\theta}=\mu^{\theta\theta}=\frac{\sin\theta}{\cos(\tilde{r}/L)},~~~~
\varepsilon^{\varphi\varphi}=\mu^{\varphi\varphi}=\frac{1}{\cos(\tilde{r}/L)\sin\theta}.
\end{equation}
where $(\tilde{r}, \theta, \varphi)$ denote the spherical
coordinates. In terms of the Cartesian coordinates
\begin{equation}\label{p2}
\varepsilon^{ij}=\mu^{ij}=\frac{1}{\cos(\tilde{r}/L)}(\delta^{ij}-(\frac{L^{2}}{\tilde{r}^{2}}\sin^{2}(\tilde{r}/L)-1)\mbox{
}\frac{x^ix^j}{ \tilde{r}^2}).
\end{equation}
The event horizon at $\tilde{r}=\pi L/2$ now becomes the boundary of
a cavity of metamaterials.

The Maxwell equations in inhomogeneous medium are
\begin{equation}\label{MX1}
 \nabla_{i}E^{i}=0,\mbox{ }\mbox{ }\mbox{ }\nabla_{i}H^{i}=0,
\end{equation}
\begin{equation}{\label{MX2}}
 \partial_{t}E^{i}-\frac{\epsilon^{ijk}}{\sqrt{\gamma}}\partial_{j}H_{k}=0,\mbox{ }\mbox{
 }\partial_{t}H^{i}+\frac{\epsilon^{ijk}}{\sqrt{\gamma}}\partial_{j}E_{k}=0,
\end{equation}
where $\gamma^{ij}$ is the optical metric related to permittivity
and permeability through
$\varepsilon^{ij}=\mu^{ij}=\sqrt{\gamma}\gamma^{ij}$, $
\gamma=\mbox{det}(\gamma_{ij})$ and $\nabla_{i}$ denotes the
covariant derivative with respect to $\gamma^{ij}$, and all the
indices are raised and lowered by $\gamma^{ij}$. To keep the
realness of the frequency of electromagnetic wave and the finiteness
of energy, the following boundary conditions are imposed
\begin{equation}\label{BC2}
 E_{\theta}|_{r=L-d}=E_{\varphi}|_{r=L-d}=0.
\end{equation}
These boundary conditions are acceptable physically, since the the
photons emitted from the center of de Sitter space will travel an
infinite amount of time to arrive at the horizon or they can never
reach there as seen by any static observer.

To solve Maxwell equations we adopt Newman-Penrose formalism
\cite{Newman:1961qr}. That is to use four null vectors reexpress the
Maxwell tensor $F_{\mu\nu}$ as
\begin{equation}\label{Fuv}
    F_{\mu\nu}=2[\phi_1(n_{[\mu}l_{\nu]}+m_{[\mu}m^{*}_{\nu]})+\phi_2l_{[\mu}m_{\nu]}+\phi_0m^*_{[\mu}n_{\nu]}]+c.c,
\end{equation}
where ``[~]" denotes the antisymmetrization, and ``c.c" means the
complex conjugate. The convention about $\phi s$ and the null
vectors is given by
\begin{eqnarray}\label{phi}
 \phi_{0}=F_{\mu\nu}l^{\mu}m^{\nu}\mbox{ },\mbox{
 }\phi_{1}=\frac{1}{2} F_{\mu\nu}(l^{\mu}n^{\nu}+m^{\ast\mu}m^{\nu})\mbox{
 },\mbox{ }\phi_{2}=F_{\mu\nu}m^{\ast\mu}n^{\nu},
\end{eqnarray}
with
\begin{eqnarray}\label{ln}
 l^{\mu}&=&(\frac{1}{1-r^{2}/L^{2}},1,0,0)\mbox{ },\mbox{ }\mbox{ }m^{\mu}=\frac{1}{\sqrt{2}r}(0,0,1,\frac{i}{\sin\theta}) \nonumber \\
n^{\mu}&=&(\frac{1}{2},-\frac{1-r^2/L^2}{2},0,0) \mbox{ },\mbox{
 }\mbox{ }
 m^{\ast\mu}=\frac{1}{\sqrt{2}r}(0,0,1,\frac{-i}{\sin\theta}).
\end{eqnarray}
To solve Maxwell equations conveniently, we have adopted a
coordinate system different from that appearing in (\ref{M}), but we
will transform back to the old coordinates when finding the typical
frequency.
 Then after some standard steps \cite{Teukolsky:1973ha}, we obtain
\begin{equation}\label{phi1}
 \phi_{1}=e^{-i w
t}Y_l^m(\theta,\varphi)R(r),
\end{equation}
where $Y_l^m(\theta,\varphi)$ is the spherical harmonic function
satisfies the following equation
\begin{eqnarray}\label{spherical}
[\frac{1}{\sin\theta}\partial_{\theta}(\sin\theta\partial_{\theta})+\frac{\partial^2_{\varphi}}{\sin^2\theta}+l(l+1)]Y_l^m(\theta,\varphi)=0,
\end{eqnarray}
and $R(r)$ satisfies
\begin{eqnarray}\label{Rr1}
[r^{2}(1-\frac{r^{2}}{L^{2}})\partial^{2}_{r}+4r(1-\frac{3r^{2}}{2L^{2}})\partial_{r}+\frac{r^{2}w^{2}}{(1-\frac{r^{2}}{L^{2}})}-\frac{6r^{2}}{L^{2}}+2-l(l+1)]R(r)=0.
\end{eqnarray}
This equation possesses two independent solutions and the one of
physical interest is
\begin{eqnarray}\label{Rr2}
R(r)=r^{-2}(1+\frac{L}{r})^{-l-1}(1+\frac{L}{r})^{-iwL/2}(1-\frac{L}{r})^{iwL/2}F(l+1,l+1+i
wL,2l+2,\frac{2r}{L+r}),
\end{eqnarray}
it is regular at $r=0$ and  goes back to flat space result $j_{l}(w
r)/r$ when $L\rightarrow \infty$. The solution of $\phi_0$ and
$\phi_2$ can also be found, but to find out $\omega$ solving
$\phi_1$ is enough.

Recall that in an empty spherical cavity the two independent
electromagnetic modes are TE and TM corresponding to $E_r=0$ and
$H_r$=0 respectively. Here the situation is similar. In the TE
modes, $E_r=0$ the electrowave is transverse. From the definition of
$\phi s$ eq.(\ref{phi}), we read $\phi_{1}=-\frac{1}{2} (E_{r}+ i
H_{r})=-\frac{i}{2}H_{r}$. Combining the following Maxwell equations
and boundary conditions (\ref{BC2})
\begin{eqnarray}\label{Hr1}
iwH_{r}=\frac{1}{r^{2}\sin\theta}(\partial_{\theta}E_{\varphi}-\partial_{\varphi}E_{\theta}),~~~
E_{\theta}|_{r=L-d}=E_{\varphi}|_{r=L-d}=0,
\end{eqnarray}
We deduce that
\begin{equation}\label{BC3}
    \phi_1|_{r=L-d}=0.
\end{equation}
Since $d\ll L$, $L-d\sim L$, the boundary conditions can be imposed
on the $r\rightarrow L$ behavior of $\phi_1$. When $r\rightarrow L$,
the radial part of $\phi_1$ has the following asymptotic form
\begin{eqnarray}\label{Rr3}
R(r)\sim\frac{\Gamma(-iwL)}{\Gamma(l+1-iwL)}(\frac{1-r/L}{1+r/L})^{iwL/2}+
c.c.
\end{eqnarray}
Then eq.(\ref{BC3}) requires that
\begin{equation}\label{solution}
    {\rm
Re}(\frac{\Gamma(-iwL)}{\Gamma(l+1-iwL)}(\frac{1-r/L}{1+r/L})^{iwL/2})|_{r=L-d}=0,
\end{equation}
In the coordinate system eq.(5), the cut-off $d$ is imposed on the
physical radial coordinate $\tilde{r}$ (\ref{M}), the cut-off on $r$
is therefore $d^2/(2L)$, the above condition amounts to
\begin{equation}\label{Condition1}
    {\rm
Re}(\frac{\Gamma(-iwL)}{\Gamma(l+1-iwL)}(\frac{1-r/L}{1+r/L})^{iwL/2})|_{r=L-d^2/2L}=0
\end{equation}
which determines value of $\omega$. For small $l$, we can solve
eq.(\ref{Condition1}) to pick out the lowest $\omega$ when
$\ln(4L^2/d^2)\gg1$ (This is guaranteed by the fact that in usual
metamaterials $d$ is nanometer, and $L$ is 1cm) and the other
corresponds to very large $n$. The results are exhibited below.

1) $l=0$. We find that (\ref{Condition1}) leads to
\begin{equation}\label{l0}
    \sin[\frac{\omega L}{2}\ln(4L^2/d^2)]=0\Rightarrow
\omega=\frac{2n\pi}{L\ln(4L^2/d^2)}, n=1,2\cdots.
\end{equation}

2) $l=1$. Then (\ref{Condition1}) implies that
\begin{equation}\label{l1}
    \omega L=\tan(\frac{\omega L}{2} \ln(4L^2/d^2)).
\end{equation}
Since $\ln(4L^2/d^2)\gg1$, the lowest $\omega\approx2\pi/L
\ln(4L^2/d^2)$.

Based on the above results, we infer that the Casimir energy from
the frequencies corresponding to small $l$ is of the order $1/L
\ln(4L^2/d^2)$ which is much smaller than $L/d^2$ predicted in
\cite{Li:2009pm}. So the typical frequency cannot be around $1/L
\ln(4L^2/d^2)$. To estimate the typical frequency, we observe that
for $l\gg1$ the Stirling formula can be used to reexpress
eq.(\ref{Condition1}) as
\begin{equation}\label{condition2}
    {\rm Re}[\Gamma(-i\omega L)(dl/2L)^{i\omega L}]=0.
\end{equation}
We see that a critical $l$ denoted by $l_{\rm c}$ emerges at $l_{\rm
c}=2L/d$ (Indeed $l_{\rm c}\gg1$, so our estimation is reasonable).
For $l\gg l_{\rm c}$, the term $(dl/2L)^{i\omega L}$ becomes highly
oscillating, and the $\omega$ satisfying (\ref{condition2})
approaches a continuous distribution whose effect is canceled by
that from the infinite background. For $l\ll l_{\rm c}$, this case
is just what we discussed before, their contribution is subleading.
Therefore the dominating contribution to the Casimir energy can only
come from $l\sim l_{\rm c}$. When $l\sim l_{\rm c}$, the
corresponding frequency is around $1/L$ since in this case the
constant appearing (\ref{condition2}) is of order 1. Go one step
further, we estimate the contribution to the Casimir energy from
frequencies corresponding to $l\sim l_{\rm c}$. For $l_{\rm
c}/2<l<3l_{\rm c}/2$, for each $l$ the degeneracy is $2l+1$,
\begin{equation}\label{EC}
    E_c\sim\sum_{l_{\rm c}/2}^{3l_{\rm c}/2}(2l+1)1/L\sim l^2_{\rm
c}/L=L/d^2.
\end{equation}
This result is the same order as we obtained in \cite{Li:2009pm} by
a different method. It is a strong support to our estimation about
the typical frequency.

In the TM modes, the magnetic wave is transverse
\begin{equation}\label{Hr3}
 H_{r}=0,~~~~~~\phi_1=-\frac{1}{2}E_r.
\end{equation}
Combined with the Gauss law and the boundary conditions (\ref{BC2})
\begin{equation}\label{con3}
    [\partial_{r}(r^{2}\sin\theta
E_{r})+\frac{\sin\theta
}{1-\frac{r^2}{L^2}}\partial_{\theta}E_{\theta}+\frac{1}{(1-\frac{r^2}{L^2})\sin\theta}\partial_{\varphi}E_{\varphi}]|_{r=L-d^2/2L}=0
\end{equation}
We deduce that
\begin{eqnarray}\label{con5}
\partial_{r}(r^{2}\phi_1)|_{r=L-d^2/2L}=\partial_{r}(r^{2}E_{r})|_{r=L-d^2/2L}=0.
\end{eqnarray}
After some simplification, this condition is transformed to
\begin{equation}\label{Condition3}
    {\rm
Im}(\frac{\Gamma(-iwL)}{\Gamma(l+1-iwL)}(\frac{1-r/L}{1+r/L})^{iwL/2})|_{r=L-d^2/2L}=0.
\end{equation}
Then the process of finding frequency $\omega$ satisfying above
condition is the same as before and we list the results as follows

1) $l=0$ the frequency is given by
\begin{equation}\label{l10}
    \cos[\frac{\omega L}{2}\ln(4L^2/d^2)]=0\Rightarrow
\omega=\frac{(2n+1)\pi}{L\ln(4L^2/d^2)}, n=0,1\cdots.
\end{equation}

2) $l=1$.
\begin{equation}\label{l20}
    \omega L=-\cot(\frac{\omega L}{2} \ln(4L^2/d^2)).
\end{equation}
For $\ln(2L/d)\gg1$, the lowest frequency is
\begin{equation}\label{l30}
    \omega\approx \pi/L\ln(4L^2/d^2).
\end{equation}

3) The typical frequency can be read from
\begin{equation}\label{condition4}
    {\rm Im}[\Gamma(-i\omega L)(dl/2L)^{i\omega L}]=0.
\end{equation}
Since the structure of eq.(\ref{condition4}) and
eq.(\ref{condition2}) is similar, the typical frequency is still of
the order $1/L$ with the critical $l_{\rm c}$ given by $2L/d$.

 We emphasize that to estimate the typical
frequency we have assumed that the contribution from $\omega$ with
large radial quantum number $n$ is suppressed exponentially. As a
check of this assumption, we present the expression of $\omega$ in
large $n$ limit. Both TE and TM modes have the following frequencies
\begin{equation}\label{fre6}
\omega\approx\frac{n\pi}{L\ln(4L^2/d^2)}.
\end{equation}
Thus for $n$ is large, $\omega$ grow with $n$ linearly and their
effects will be suppressed by the black body factor.

To end this section, we propose that the measurable quantity for
such experiments would be the Casimir force. From eq.(\ref{EC}), we
get
\begin{equation}\label{force}
F_{C}\sim -\frac{1}{d^{2}}.
\end{equation}
It is remarkable that this attractive force is relatively large
compare with the usual Casimir force. Thus, it is more easily to
measure the new Casimir force for a cavity with small cutoff. We
also notice that the unusually large Casimir force is related to the
permittivity and the permeability at $\tilde{r}=\frac{\pi}{2}(L-d)$
(\ref{M})
\begin{equation}\label{M1}
\varepsilon=\mu\sim \frac{L}{d}.
\end{equation}
where $d$ is some microcosmic cutoff to keep the permittivity and
the permeability finite but relatively large.
\section{Conclusion}

We estimate the typical frequency of the vacuum fluctuations in
metamaterials mimicking de Sitter and find it proportional to $1/L$,
the size of the cavity. Assuming  $d$ be 1 nanometer, and $L$ be
1cm, then the typical wavelength is about 1cm.

With our estimation of the typical frequency and the typical angular
quantum number, we also have an intuitive understanding of our
Casimir energy formula.

We hope that one day the metamaterials suggested by us can be made
with appropriate size and effective for the corresponding typical
frequency, then the predicted brand new Casimir force can be
measured, this experiment is important for study cosmology in
laboratory.

\section*{Acknowledgments}

We would like to thank Prof. Mo Lin Ge for informing us of the
exciting developments in the field of electromagnetic cloaking and
metamaterials. This work was supported by the NSFC grant
No.10535060/A050207, a NSFC group grant No.10821504 and Ministry of
Science and Technology 973 program under grant No.2007CB815401.

\end{document}